\documentclass[10pt]{article}
\usepackage[OE]{express}
\usepackage{amsmath}

\usepackage{graphicx}

\usepackage{graphicx}
\usepackage{epstopdf}
\usepackage{braket}
\usepackage{float}
\usepackage{color}
\usepackage{placeins}
\usepackage{tikz}
\usetikzlibrary{arrows.meta}

\begin{document}

\newcommand{\apar}{\alpha_\parallel}
\newcommand{\aperp}{\alpha_\bot}
\newcommand{\tj}[6]{
\left(
\begin{array}{ccc}
#1 & #2 & #3 \\
#4 & #5 & #6 
\end{array}
\right)
}
\title{Elliptical polarization for molecular Stark shift compensation in deep optical traps}

\author{Till Rosenband,\authormark{1,*} David D. Grimes,\authormark{2,1,3} and \\Kang-Kuen Ni\authormark{2,1,3}}

\address{\authormark{1}Department of Physics, Harvard University, Cambridge, MA 02138, USA\\
\authormark{2}Department of Chemistry and Chemical Biology, Harvard University, Cambridge, MA 02138, USA\\
\authormark{3}Harvard-MIT Center for Ultracold Atoms, Cambridge, MA 02138, USA}

\email{\authormark{*}rosenband@fas.harvard.edu} 



\begin{abstract}
In optical dipole traps, the excited rotational states of a molecule may experience a very different light shift than the ground state.  For particles with two polarizability components (parallel and perpendicular), such as linear $^1\Sigma$ molecules, the differential shift can be nulled by choice of elliptical polarization.  When one component of the polarization vector is $\pm i\sqrt{2}$ times the orthogonal component, the light shift for a sublevel of excited rotational states approaches that of the ground state at high optical intensity.  In this case, fluctuating trap intensity need not limit coherence between ground and excited rotational states.
\end{abstract}
\\


\section{Introduction}
Optical dipole traps are the reservoirs that enable many experiments with cold isolated atoms and molecules.  In most cases, particles are attracted to the intensity maximum of a focused optical beam or a standing wave due to their polarizability.  These traps are usefully characterized by their trap depth, the light shift at the intensity maximum, which can be expressed in frequency units.  Typical trap depths range over many orders of magnitude, from below $1~$kHz to above $10~$MHz.  The light not only traps the particle, but also perturbs the energy eigenstates and causes relative shifts between energy levels. The perturbation is significant for molecules, where the intrinsic anisotropy of polarizability can cause rotationally excited states to have energy shifts with respect to the ground state that are large fractions of the trap depth.

A ``magic'' angle~\cite{Cagnac1968,Happer1970,Budker2004,Kotochigova2010} has been found between the linear polarization of light and the quantization axis, which causes energy states of atoms and excited rotational states of $^1\Sigma$ molecules to have the same light shift as the ground state.  For molecules, the ``magic'' angle changes and applies to only one sublevel of the first rotationally excited state when the light shift becomes significant compared to the splittings between sublevels~\cite{Neyenhuis2012}, and the nulling angle depends on the intensity, electric field, magnetic field, and internal hyperfine couplings~\cite{Gregory2017}. In atoms, elliptical polarizations have been found that eliminate first-order light shifts between atomic hyperfine states~\cite{HKim2013} as well as second-order light shifts of optical transitions~\cite{Taichenachev2006}.

Sometimes it is desirable to confine $^1\Sigma$ molecules in a trap whose light shift is larger than the sublevel splittings (e.g. due to hyperfine and Zeeman coupling), but still small compared to the energy difference between ground and second excited rotational states.  For the purpose of this paper, we use the term ``deep trap'' to refer such a situation.  Then there is no ``magic'' angle of linear polarization, because the molecule aligns itself to the polarization axis and the polarization direction becomes the dominant quantization axis.  However as shown below, a particular elliptical polarization causes one sublevel of rotationally excited states to have the same light shift as the ground state.

\section{Elliptical polarization}
Deep optical traps typically operate at frequencies far below electronic resonances to reduce photon scattering.  We assume that the optical frequency has a large detuning with respect to the rotational and vibrational structure of electronic excited states.  In this case, the polarizability tensor of $^1\Sigma$  molecules is symmetric and vector light shifts are absent~\cite{Berestetskii}. For linear molecules~\cite{James1959} and other symmetry groups~\cite{Bonin1997}, the molecular-frame polarizability tensor has the simple form 
\begin{equation}
\tilde{A}=\left(
\begin{array}{ccc}
\aperp &   0 & 0 \\
0   & \aperp & 0 \\
0   &   0 & \apar
\end{array}
\right),
\end{equation}
leading to a light shift of $-E^\dagger \tilde{A} E/2$~\cite{Happer1970,Friedrich1995,Stapelfeldt2003} where $\apar$ and $\aperp$ are real parallel and perpendicular polarizabilities, and $E$ is the electric field vector in the molecular frame.  This expression is valid for rapidly oscillating fields $Re(E e^{i\omega t}\sqrt{2})$ where $E=\epsilon E_0$, $\epsilon$ is the complex unit polarization vector, and $E_0$ is the RMS electric field amplitude.  To convert to the laboratory frame for a molecule whose orientation lies along the spherical coordinates $\theta,\phi$ we apply the rotations
\begin{equation}
\tilde{R}_x=\left(
\begin{array}{ccc}
1 &   0 & 0 \\
0   & \cos{\theta} & -\sin{\theta} \\
0   & \sin{\theta} & \cos{\theta}
\end{array}
\right),
\tilde{R}_z=\left(
\begin{array}{ccc}
\cos{\phi} & -\sin{\phi} & 0\\
\sin{\phi} & \cos{\phi} & 0\\
0 & 0 & 1
\end{array}
\right)
\end{equation}
\clearpage
\begin{equation}
\tilde{\alpha}(\theta,\phi)=\tilde{R}_z^\dagger \tilde{R}_x^\dagger \tilde{A} \tilde{R}_x \tilde{R}_z.
\end{equation}

The light shift (quadratic Stark shift) for laboratory-frame electric fields causes an effective potential energy for the molecule that is a function of orientation $\theta,\phi$:
\begin{equation}
\label{eqStark}
V(\theta,\phi)=-\frac{1}{2}E_0^2\epsilon^\dagger \tilde{\alpha}(\theta,\phi) \epsilon
\end{equation}
In this treatment, we neglect the linear interaction between the electric field and the permanent molecular dipole moment, because the optical field is far above the rotation frequency.
  
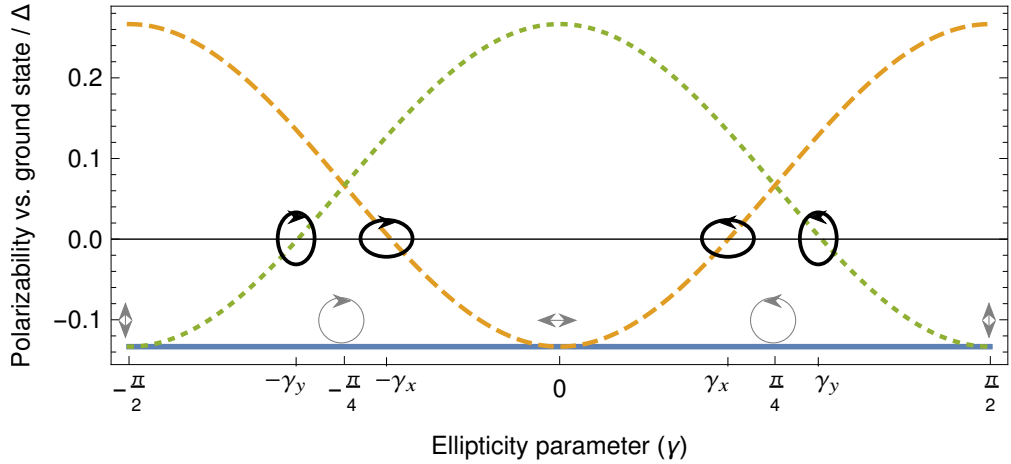
\begin{figure}
  \centering  
\begin{tikzpicture}
 \node[anchor=south west,inner sep=0] (image) at (0,0) {
    \includegraphics[width=\linewidth]{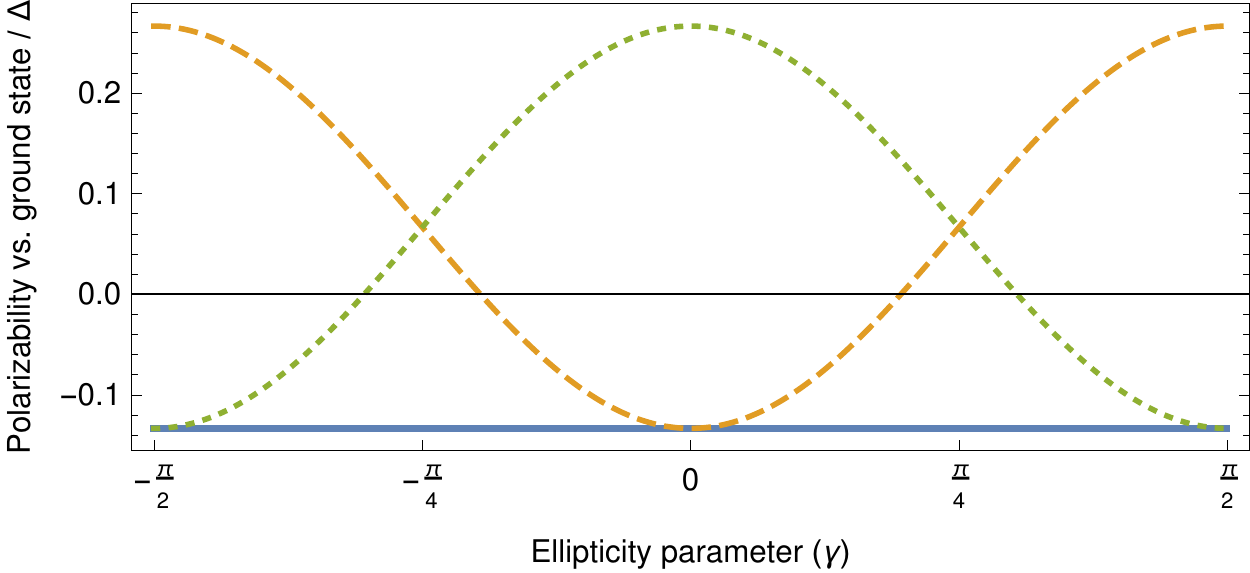}};


\begin{scope}[x={(image.south east)},y={(image.north west)}]
\draw[gray](0.335,0.31) circle(3mm);
\draw[gray] [-{Stealth[scale=1.5]}] (0.335,0.352) -- (0.345,0.352);
\draw[gray](0.765,0.31) circle(3mm);
\draw[gray] [-{Stealth[scale=1.5]}] (0.765,0.352) -- (0.755,0.352);

\draw[line width=0.5mm](0.72,0.488) ellipse(3.46mm and 2.45mm); 
\draw[] [-{Stealth[scale=1.5]}] (0.72,0.523) -- (0.71,0.523);
 
\draw[line width=0.5mm](0.81,0.488) ellipse(2.45mm and 3.46mm);
\draw[] [-{Stealth[scale=1.5]}] (0.81,0.534) -- (0.8,0.534);

\draw[line width=0.5mm](0.38,0.488) ellipse(3.46mm and 2.45mm);
\draw[] [-{Stealth[scale=1.5]}] (0.38,0.523) -- (0.39,0.523);

\draw[line width=0.5mm](0.29,0.488) ellipse(2.45mm and 3.46mm);
\draw[] [-{Stealth[scale=1.5]}] (0.29,0.534) -- (0.3,0.534);

\draw[] [-] (0.29,0.23) -- (0.29,0.21) ;
\draw[] (0.28,0.21) node[anchor=north] {\bf $-\gamma_y$};

\draw[] [-] (0.38,0.23) -- (0.38,0.21) ;
\draw[] (0.39,0.21) node[anchor=north] {\bf $-\gamma_x$};

\draw[] [-] (0.72,0.23) -- (0.72,0.21) ;
\draw[] (0.71,0.21) node[anchor=north] {\bf $\gamma_x$};

\draw[] [-] (0.81,0.23) -- (0.81,0.21) ;
\draw[] (0.82,0.21) node[anchor=north] {\bf $\gamma_y$};

\draw[gray] [<->,>={Stealth[scale=1.5]}] (0.53,0.31) -- (0.57,0.31);
\draw[gray] [<->,>={Stealth[scale=1.5]}] (0.12,0.27) -- (0.12,0.35);
\draw[gray] [<->,>={Stealth[scale=1.5]}] (0.98,0.27) -- (0.98,0.35);
\end{scope}

\end{tikzpicture}
\caption{Eigenvalues of $\hat{\beta}(\gamma)$, as defined in Eq.~(\ref{eqBetaMatrix}), in units of $\Delta$ vs. ellipticity parameter $\gamma$.  Linear polarization corresponds to $\gamma=0$ or $\pm\pi/2$ where two values are negative and one positive. Circular polarization corresponds to $\gamma=\pm\pi/4$ where two values are positive and one negative.  Between these extremes are zero-crossings where one $J=1$ state has no differential light shift with respect to the $J=0$ state. Polarizations are depicted with horizontal $\hat{x}$ and vertical $\hat{y}$ directions.}
\label{figCirc}
\end{figure}

We define a variable elliptical polarization in the $\hat{x}\hat{y}$ plane as $\epsilon(\gamma)=\hat{x}\cos{\gamma}+i\hat{y}\sin{\gamma}$.  The differential polarizability for light of this polarization with respect to the ground state is $\beta(\gamma,\theta,\phi)=\epsilon(\gamma)^\dagger\tilde{\alpha}(\theta,\phi)\epsilon(\gamma) - \alpha_{0}$, where the ground state polarizability is $\alpha_0=(\apar+2\aperp)/3$.  Via trigonometric substitutions, $\beta$ can be written in terms of spherical harmonics $Y_{J,m}(\theta,\phi)$ as
\begin{eqnarray}
\beta(\gamma,\theta,\phi)&=&
 \Delta\left(-\frac{1}{3}+\frac{1}{2}\sin^2{\theta}-
 \frac{1}{2}\cos{(2\gamma)}\cos{(2\phi)}\sin^2{\theta}\right) \\
 &=&-\Delta\sqrt{\frac{2\pi}{15}}\sum_{m=-2}^{2}{c_m Y_{2,m}(\theta,\phi)}
 \label{eqYlm}
\end{eqnarray}
where $\Delta=\apar-\aperp$, $c_0=\sqrt{2/3}$, $c_{\pm1}=0$, and $c_{\pm2}=\cos{(2\gamma)}$.  Equation~(\ref{eqYlm}) is advantageous, because the basis state for rotation of the internuclear axis with  angular momentum $J$ and projection $m$ along the laboratory $z$ axis is the spherical harmonic wavefunction $Y_{J,m}(\theta,\phi)$.  For these states, the matrix elements of $\hat{\beta}(\gamma)$ are sums of integrals of products of three spherical harmonics that can be written in terms of Wigner-3j symbols~\cite{Zare1988}, where elements with $|m'-m|>2$ are zero:
\begin{eqnarray}
\label{eqMEbeta}
\nonumber
\braket{J',m'|\hat{\beta}(\gamma)|J,m} & \equiv & \int Y_{J',m'}(\theta,\phi)^* Y_{J,m}(\theta,\phi)\beta(\gamma,\theta,\phi)d\Omega \\ \nonumber
&=& -\Delta (-1)^{m'}\sqrt{\frac{(2J'+1)(2J+1)}{6}}
   \tj{2}{J}{J'}{0}{0}{0} \times \\
  & & c_{m'-m} \tj{2}{J}{J'}{m'-m}{m}{-m'}
\end{eqnarray}

A large light shift can dominate the molecular hyperfine and Zeeman Hamiltonians~\cite{Aldegunde2017} of $^1\Sigma$ molecules while still being small compared to the energy splitting between states with different $J$. In this case, the energy eigenstates for each rotational level $J$ approximate the eigenstates of $\hat{\beta}(\gamma)$.  An eigenvalue of zero corresponds to states whose light shift approaches the ground state light shift. For $J=1$, we find in the basis $m=-1,0,1$
\begin{equation}
\label{eqBetaMatrix}
\hat{\beta}(\gamma)=\frac{\Delta}{15}
\left(
\begin{array}{ccc}
 1 & 0 & 3 \cos (2 \gamma ) \\
 0 & -2 & 0 \\
 3 \cos (2 \gamma ) & 0 & 1 \\
\end{array}
\right).
\end{equation}

As shown (Fig.~\ref{figCirc}), the three eigenvalues are
$(\Delta/15)\{-2,1+3\cos{(2\gamma)},1-3\cos{(2\gamma)}\}$.  When $\cos{(2\gamma)}=\pm1/3$, two eigenvalues are $\pm 2\Delta/15$ and the other equals zero, making its $J=1$ eigenstate have a constant energy offset from the $J=0$ state, even in the face of  trap intensity fluctuations.  This occurs for $\pm\gamma_x$ and $\pm\gamma_y$ where $\gamma_x\approx 35.26^\circ$ and $\gamma_y\approx 54.74^\circ$ are complementary angles and the sign gives the sense of circularity. The ``magic'' angle for linear polarization in weaker traps is identical to $\gamma_y$.  Note that the parameterization for $\epsilon$ defined the ellipse's major axis to lie along $\hat{x}$ or $\hat{y}$, but more generally it could be arbitrarily oriented.  For higher rotation states $J'=J>1$, one can verify from the matrices defined by Eq.~(\ref{eqMEbeta}) that the same polarization ellipticity causes one eigenstate to have zero differential light shift with respect to $J=0$, if physical effects like centrifugal forces, and the frequency dependence of polarizability are neglected~\cite{Zwierlein2018APC,RotationXYZ}.

For the choice $\gamma_x$, the polarization vector is $\epsilon=\hat{x}\sqrt{2/3}+i\hat{y}/\sqrt{3}$, and the sensitivity to imperfect polarization can be found from a series expansion in $\gamma$ of the polarizability difference.  To first order, the difference for the nulled $J=1$ state changes by $(4\sqrt{2}/15)\delta\Delta$ where $\delta$ is the change in $\gamma$.  Note that the results are true for all values $\apar$ and $\aperp$, but the range of deep traps for the purpose of this calculation depends on their difference $\Delta$.  

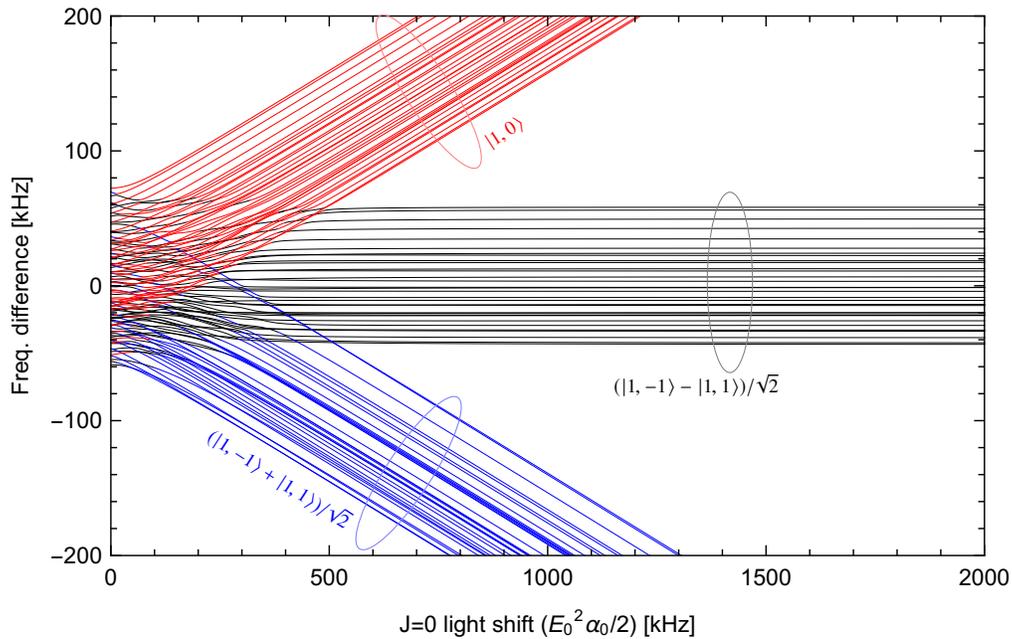
\begin{figure}[t]
\begin{tikzpicture}
 \node[anchor=south west,inner sep=0] (image) at (0,0) {
  \includegraphics[width=\linewidth]{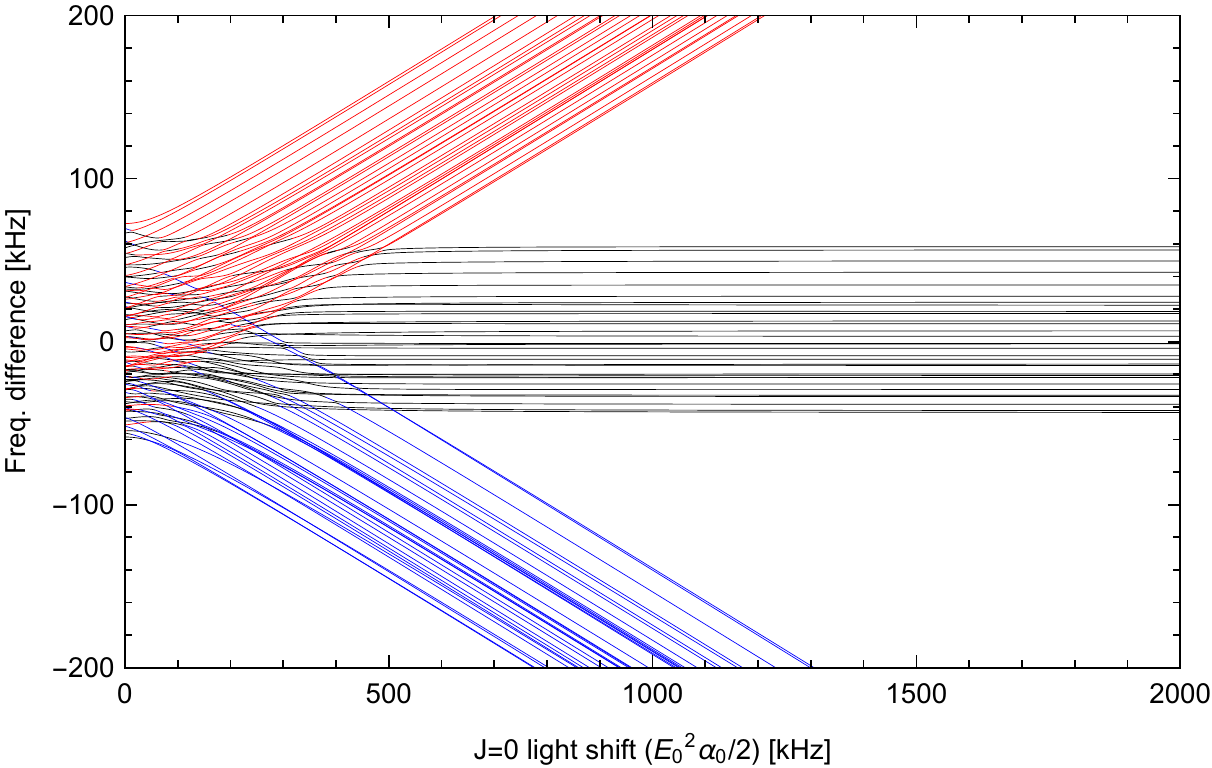}};

\begin{scope}[x={(image.south east)},y={(image.north west)}]
\draw[blue!50, rotate around={-33:(0.4,0.26)}](0.4,0.26) ellipse(3mm and 12mm);
\draw[black!50](0.72,0.56) ellipse(3mm and 12mm);
\draw[red!50, rotate around={33:(0.42,0.86)}](0.42,0.86) ellipse(3mm and 12mm);
\end{scope}
\end{tikzpicture}
  \put(-305,75){\rotatebox{-33}{\textcolor{blue}{\scriptsize $(\ket{1,-1}+\ket{1,1})/\sqrt{2}$}}}
  \put(-200,185){\rotatebox{33}{\textcolor{red}{\scriptsize $\ket{1,0}$}}}
  \put(-150,93){\rotatebox{0}{\scriptsize $(\ket{1,-1}-\ket{1,1})/\sqrt{2}$}}
\caption{Difference of energy eigenvalues between $J=1$ and $J=0$ as a function of $J=0$ light shift for an example molecule ($NaCs$ with 10 Gauss magnetic field along $\hat{z}$) where the offset of 3.48~GHz due to rotational energy is removed.  The polarizability of the rotation state $(\ket{1,-1}-\ket{1,1})/\sqrt{2}$, approaches that of the ground state for high intensity when the polarization vector is $\epsilon=\hat{x}\sqrt{2/3}+i\hat{y}/\sqrt{3}$, as described in the text.  We add the molecular Hamiltonian $H_0$~\cite{Aldegunde2017} to the light shift of Eq.~(\ref{eqMEbeta}), to make the 32 molecular hyperfine states visible.  Tensor coupling between nuclear magnetic dipoles is neglected due to smallness.  The hyperfine coupling and Zeeman energies of $H_0$ must be small compared to $2E_0^2\Delta/15$ for the deep-trap limit to apply.  A polarizability anisotropy of $\aperp=\apar/4$ is used, corresponding to $NaCs$ at a wavelength of 1064~nm~\cite{Vexiau2017}, but the exact values do not affect the main result. The eigenstates for deep traps are labeled in terms of superpositions of $\ket{J,m}$ basis states.  We have neglected $J=2$ states in this calculation, and their energy difference with the ground state sets an upper bound for the range of trap depths where the results are valid.}
\label{figEP}
\end{figure}

Figure~\ref{figEP} shows an example where elliptically polarized light is applied to an NaCs bialkali molecule and the effects of hyperfine and Zeeman structure are included~\cite{Aldegunde2017}. Note that the majority of bialkali molecules have larger hyperfine splittings and larger light shifts are needed to reach the ``deep trap'' regime described here.  For molecules where the coupling between nuclear spin and molecule rotation is smaller, a reduced light shift can suffice. Molecular states with electron spin (e.g. $^2\Sigma$ and $^3\Sigma$) have additional spin-rotation coupling that must be overcome for the light shift to dominate.

\section{Conclusion}
Experiments that explore dipole-dipole interactions between nearby $^1\Sigma$ molecules~\cite{Yan2013} utilize rotation states with $|J-J'|=1$, such as $J=0$ and $J'=1$.  These experiments benefit from a consistent interaction strength, which requires small position variance, because the strength scales as the inverse cube of the intermolecular distance.  A tightly-confining optical dipole trap with large intensity helps reduce the variance, and the elliptical polarization described above nulls the associated differential light shifts.  Reduction of differential light shifts is also important for precision measurements.  It was previously shown that it is possible to cancel these shifts at low intensity by choice of linear polarization angle~\cite{Kotochigova2010,Neyenhuis2012,Gregory2017}, and we have shown that elliptical polarization can serve the same purpose at high intensity.  The use of elliptical polarization may have practical advantages, because amplitude fluctuations have a vanishing effect in the high-intensity limit, as long as light shifts are well below the spacing of rotational levels. In addition, experimental drifts of the polarization parameter $\gamma$ are likely to be slow thermal drifts, whose phase errors can be unwound by spin echo techniques~\cite{Viola1998}.

\section*{Funding}
National Science Foundation (NSF) (PHY-1125846); David and Lucile Packard Foundation (2016-65138).

\section*{Acknowledgments}
We thank Rosario~Gonz\'alez-F\'erez for valuable comments.  DDG is supported by a fellowship from the Max Planck Harvard Quantum Optics Center.


\begin{thebibliography}{99}
\bibitem{RotationXYZ}
Polarization with ellipticity $\epsilon=i(\hat{x}~\mathrm{or}~\hat{y})\sqrt{2/3} \pm \hat{z}\sqrt{1/3}$ in the $xz$ or $yz$ plane rather than $xy$ causes the diagonal elements of $\hat{\beta}$ to be zero in Eq.~\ref{eqMEbeta}, making it apparent that for all $J$, $\hat{\beta}$ is singular with a zero eigenvalue.
\end{thebibliography}
\end{document}